\documentstyle[eqsecnum,prd,aps,epsf]{revtex}
\def\abstract#1{\vskip 7mm
\begin{center}{\large Abstract}\par \smallskip \begin{minipage}[c]{12cm}
\small #1
\end{minipage}
\end{center}
}
\def\title#1{\begin{center}{\Large\bf #1}\end{center}} 
\def\author#1{\vskip 5mm \begin{center}{#1}\end{center}} 
\def\address#1{\begin{flushleft}{\it #1}\end{flushleft}}

\def\non{\nonumber}
\def\be{\begin{equation}}
\def\ee{\end{equation}}     
\def\bea{\begin{eqnarray}}
\def\eea{\end{eqnarray}}
\newcommand{\vect}[1]{\!\!\!\mbox{ \boldmath $#1$}}

\def\Ucl{U_{\mbox{c-low}}(\epsilon)}
\def\Uch{U_{\mbox{c-high}}(\epsilon)}
\def\G{{$\cal G$} }
\def\C{{$\cal C$} }
\def\I{{$\cal I$} }
\def\core{{\it core} }
\def\halo{{\it halo} }
\def\gas{{\it gas} }
\def\HALO{{\it HALO} }

\def\lc{\lambda_{\mbox{\it core}}}
\def\lh{\lambda_{\mbox{\it halo}}}
\def\lg{\lambda_{\mbox{\it gas}}}

\def\lck{\lambda^k_{\mbox{\it core}}}
\def\lhk{\lambda^k_{\mbox{\it halo}}}
\def\lgk{\lambda^k_{\mbox{\it gas}}}

\def\tauu{t_{R}}
\def\tauf{t_{\xi}}
\def\tauc{t_{E}}
\def\tautc{\tau_{E}}
\def\tautci{\tau_{Ei}}
\def\tautcz{\tau_{E1}}
\def\tautcf{\tau_{E1}}
\def\tautcs{\tau_{E2}}
\def\tautf{\tau_{\xi}}
\def\taurec{\tau_{\mbox{rec}}}

\begin{document}

\title{%
Origin of 
scaling structure and non-gaussian velocity
distribution in  self-gravitating ring model} 
\author{%
Yasuhide Sota$^{1,3}$
\footnote[1]{E-mail:sota@skyrose.phys.ocha.ac.jp},
Osamu Iguchi$^1$
\footnote[5]{E-mail:osamu@phys.ocha.ac.jp},
Masahiro Morikawa$^{1}$
\footnote[2]{E-mail:hiro@phys.ocha.ac.jp}, \\[.5em]
Takayuki Tatekawa$^2$
\footnote[6]{E-mail:tatekawa@gravity.phys.waseda.ac.jp} 
,and
Kei-ichi Maeda$^{2,3,4}$
\footnote[3]{E-mail:maeda@mse.waseda.ac.jp}\\~ } 
\address{%
$^1$ Department of Physics, Ochanomizu University, 
2-1-1 Ohtuka, Bunkyo, Tokyo,112-8610 Japan\\[.3em]
$^2$ Department of  Physics, 
Waseda University,   Shinjuku, Tokyo 169-8555, Japan\\[.3em]
$^3$ Advanced Research
Institute for Science and Engineering,  Waseda University,\\~~~ Shinjuku,
Tokyo 169-8555, Japan\\[.3em]
$^4$ Advanced Institute for Complex Systems, 
Waseda University, Shinjuku, Tokyo 169-8555, Japan}

\abstract{
Fractal structures and non-Gaussian velocity distributions 
are characteristic properties commonly observed 
in virialized self-gravitating systems 
such as galaxies and interstellar molecular clouds. 
We study the origin of these properties 
using a one-dimensional ring model 
which we newly propose in this paper. 
In this simple model, $N$ particles are moving, 
on a circular ring fixed in three dimensional space,  
with mutual interaction of gravity. 
This model is suitable for the accurate symplectic integration method 
by which we argue the phase transition in this system.  
Especially, in between the extended-phase and the collapsed-phase, 
we find an interesting phase (\halo-phase) 
which has negative specific heat at the intermediate energy scale.   
Moreover in this phase, there appear scaling properties and  
non-thermal and non-Gaussian velocity distributions. 
In contrast, 
these peculiar properties are never observed 
in other \gas and \core phases. 
Particles in each phase have a typical time scale of motions 
determined by the cutoff length $\xi$, 
the ring radius $R$, and the total energy $E$. 
Thus all relaxation patterns of the system are determined 
by these three time scales. 
}

\section{Introduction}

Many astrophysical objects in our universe consist of 
mutually interacting elements through gravity. 
If they are almost isolated systems, 
they are called self-gravitating system (SGS). 
For example, 
galaxies, clusters of galaxies, globular clusters, and molecular clouds 
are thought to be typical SGSs. 
Their statistical properties are often characterized 
by non-Gaussian velocity (or pair-wise velocity) 
distributions\cite{Zurek94}, 
fractal structures\cite{Pietro98}, 
and the scaling relation 
between the mass density and the system size\cite{DeVaucouleurs}. 
Most of these objects are thought to be 
gravitationally virialized. 
Therefore pure gravitational force seems to play an essential role 
in characterizing the above statistical properties of SGS independently of initial conditions.  

There have been some theoretical approaches 
to explain the fractal structures in SGS 
from the viewpoint of criticality and phase transition 
in gravo-thermo dynamics\cite{de Vega96}. 
Strictly speaking, the ordinary SGSs in three-dimension (3-D) cannot attain 
genuine stable equilibrium, 
because the gravitational force does not vanish 
at long distances (IR-divergence) 
and diverges at short distances (UV-divergence). 
These properties of gravity cause gravothermal catastrophe 
in a self-gravitating gas system enclosed by a solid adiabatic wall. 
In fact, the isothermal sphere is not always stable, 
since the entropy does not necessarily take the local maximum 
for this configuration\cite{Antonov62,Lynden-Bell68,Hachisu78}. 
Therefore the introduction of a small-scale cutoff 
as well as a large-scale cutoff is inevitable 
for such unstable systems to attain the final equilibrium. 
Though the introduction of the cutoff prevents 
the gravothermal catastrophe, 
3-D gravitational system has a phase 
with negative specific heat when treated 
in the microcanonical ensemble\cite{Aronson72}. 
Even in the canonical ensemble, this system is highly unstable. 
When the temperature decreases, 
the system experiences the violent first-order phase transition 
from the gas phase into the cluster phase. 
According to these arguments, 
no stable equilibrium states are theoretically expected 
in the 3-D gravitational system with or without cut-off in any ensemble.  

On the other hand, in the real world, we do not need the stable equilibrium states for describing SGS; 
meta-stable equilibrium states do actually appear in the dynamical 
description such as collisionless Boltzmann equation. 
The finite lifetime of such structures is sufficient 
to explain the present structures of SGS, 
even if they are expected to evolve further into 
different quasi-stable states through two-body relaxation.   

The relaxation process of SGS has been mainly discussed
in the one-dimensional gravitational sheet model (1DS) 
\cite{Camm50,Tsuchiya96}.
In this model, 
many parallel sheets interact with each other 
through constant force which never decays at distant places. 
Though the interaction is long-ranged, 
no phase transition occurs in 1DS. 
Thermodynamics of 1DS is exactly solved and actually, 
in numerical calculations, 
the system reaches thermal equilibrium 
long after it attains virial equilibrium. 
In this model, 
the virial condition gives the relation 
$2\left< K \right> =\left< V \right>$ 
between the time averaged kinetic energy $\left< K \right>$ 
and the potential energy $\left< V \right>$. 
Therefore, contrary to the 3-D SGS, 
specific heat of 1DS is always positive. 
Thus the relaxation process in 1DS would be quite different from that in 3-D SGS.  
Another well-known one-dimensional model 
which has long-range force is the Hamilton Mean Field (HMF) model, 
in which phase transition does occur\cite{Antoni95}. 
There have been a lot of studies on the relaxation process of HMF. 
Actually, L\'evy type jumping motion of constituent particles 
and super-diffusion process have been revealed in HMF\cite{Latora99}. 
In this HMF, though the specific heat becomes negative in the quasi-stable state \cite{Latora01}, 
it remains positive in the thermal equilibrium state and 
the phase transition turns out to be the second-order.
Despite some common properties of the models, 
the interaction form in HMF is also quite different from 
that of real 3-D gravitational systems.

Several other simple models have been proposed\cite{padmanabhan90} 
in order to characterize the 3-D SGS much faithfully. 
For example, 
the cell model and its extended version are the simplest models 
which show phase transition. 
In these models, 
the pair interaction potential changes its value only 
at the cell boundary\cite{Thirring71,Compagner89}. 
At low temperatures, most particles are trapped within several clusters, 
and at higher temperatures, these clusters melt 
and the particles can move freely. 
However, 
it seems difficult to examine 
the relaxation process of this system numerically, 
since the interaction is not efficient to cause sufficient relaxation; 
particles have no interaction within each cell. 
There are also several numerical analysis 
with the different type interactions 
including the lower and higher cutoffs\cite{Compagner89,Aizawa00}. 
However, 
the value of the lower cutoff used in realistic numerical calculations 
of these models seems to be too large 
to extract the intrinsic properties of realistic 3-D SGS.

Thus it seems indispensable for us to have a model 
which faithfully reflects the characteristic properties of 3-D SGSs; 
the model has negative specific heat and shows the phase transition 
similar to 3-D SGSs, and is capable for us to analyze quasi-equilibrium states 
which would be realized before the system reaches the complete equilibrium. 

In this paper, therefore, 
we first propose a new model (Self Gravitating Ring: SGR) which 
a) has negative specific heat, 
b) shows phase transition representing 3-D SGS, and 
c) is numerically tractable.  
This model consists of $N$-particles, 
which can freely move on a ring with a fixed radius, 
mutually interacting through 3-D gravity.  
An excellent point of this model is 
that the force is genuine 3-D gravity while the calculation is 
essentially one-dimensional. 
Moreover, 
the Hamiltonian permits the accurate symplectic integration method 
by which we can analyze the nature of quasi-equilibrium states 
and phase transitions within very long time scales in this model.   

In section \ref{sec:SGR}, we introduce the SGR model, 
and in section \ref{sec:class} identify 
three quasi-equilibrium states including the state with 
negative specific heat. 
In section \ref{sec:motion}, 
we analyze the particle motions from a statistical point of view, 
and then, we study the relaxation process toward 
the thermodynamical equilibrium in section \ref{sec:relax},
and the scaling structure in section \ref{sec:scaling}. 
Finally, we discuss our results in section \ref{sec:con}.

\section{Self-gravitating ring model}
\label{sec:SGR}

In this section, 
we introduce the self-gravitating ring (SGR) model, 
in which particles interact with each other through genuine 3-D gravity 
while the particle motion is constrained on a 1-D ring. 
By utilizing this model, 
we study the gravitational phase transition, 
which was not feasible in the 1DS model\cite{Tsuchiya96}. 
The lack of phase transition in 1DS is related with the fact that 
the potential of 1DS increases 
linearly without bound, 
and therefore there is no characteristic energy scale 
required for a particle to escape from the cluster.   
On the other hand, 
we have the characteristic energy scale to bind particles in our SGR model.  

\begin{figure}[htbp] \begin{center} \leavevmode
\epsfysize=5.0cm
\epsfbox{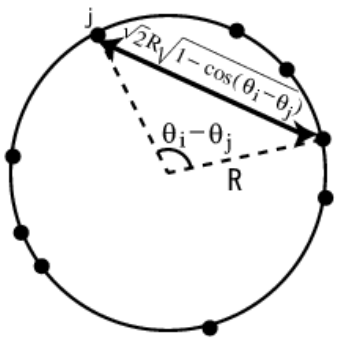}
\caption{
The SGR model with a fixed radius $R$. 
The particle locations are specified by the angles measured from a fixed direction.  
A pair of particles at $\theta_i$ and at $\theta_j$ interact with each other 
through the inverse-square-law 3-D gravitational force; the distance is 
measured by the straight line in the picture and not by $R |\theta_i-\theta_j|$. 
}
\label{fig-ring} \end{center} \end{figure}

We consider a system of self-gravitating $N$ particles with mass $m$, 
whose motions are smoothly constrained on a circular ring 
with a fixed radius $R$ without friction (Fig.\ref{fig-ring}). 
Each pair of particles interact with each other 
through 3-D gravitational force. 
The distance between a pair is measured 
by the length of the straight line combining the pair 
and not by the minor arc of the ring.  

The Hamiltonian of this system becomes
\bea
 && H_{\rm p} =
    \frac{1}{2 m R^2}\sum_{i=1}^N P_i^2
    -\sum_{i<j}^N\frac{G m^2}
    {\sqrt{2}R\sqrt{1-\cos(\theta_{i}-\theta_j )+\epsilon}}.
\label{canoeq}
\eea
The position of $i$-th particle is fully described 
by the angular variable $\theta_i$ 
as $\vect{r}_i=(R\sin\theta_i , R\cos\theta_i)$.  
The momentum conjugate to $\theta_i$ is given 
by $P_i = m R^2 d \theta_i/d t$. 
The UV-cutoff parameter $\epsilon$ truncates 
the diverging gravitational force 
at around the distance $\xi \equiv \sqrt{2\epsilon} R$.

We first introduce three dynamical time-scales which are apparent 
in the above Hamiltonian. 
They are parameterized by the ring radius $R$, 
the cutoff scale $\xi$, and the total energy $E$ of the system. 

When the system is almost uniformly filled by moving particles, 
we have the longest dynamical time $\tauu$ defined by
\be
\tauu \equiv \sqrt{R^3 \over GNm}.  
\ee
During this time $\tauu $, 
a typical particle goes around the ring once. 
Therefore, 
the time-scale for the whole system to attain thermodynamical equilibrium, 
if any, is at least larger than this time scale. 

When all the particles collapse completely into a core, 
we find the shortest dynamical time-scale $\tauf$ defined as
\be
\tauf \equiv \sqrt{{\xi^3 \over G N m}}.
\label{timefree}
\ee
During this time $\tauf$, 
a typical particle bound to the core oscillates once.
\footnote{
In the next section, we classify all the particles into three species,
\core, \halo, and \gas.  
The typical time scales 
$\tauu$ and $\tauf$ respectively characterize the \gas and the \core species.
}

There is an another time-scale in between the above two extreme time scales. 
Suppose the system is stably confined in a region with
a scale $r$ which satisfies
\bea
\xi<<r<<R.
\label{xi<<r<<R}
\eea
For this scale $r$, 
the ring can be approximated as an infinite straight line 
and the cutoff $\xi$ can be neglected. 
Then the leading term of the denominator of the potential term 
in the Hamiltonian (\ref{canoeq}) becomes $|\theta_{i}-\theta_j|$ and 
\bea
 && H_p \approx
    \frac{1}{2 m R^2}\sum_{i=1}^{N} P_i^2
    -\sum_{i<j}^{N}\frac{G m^2}{R |\theta_{i}-\theta_j |}.
\label{canoeq-app}
\eea
In this case, 
the potential term $V_{\rm p}$ satisfies Euler's theorem 
for homogeneous function,
\be
\sum_{i=1}^{N} \theta_i
 \frac{\partial V_{\rm p}}{\partial \theta_i}=-V_{\rm p},
\label{euler}
\ee
and the ordinary virial condition holds
\bea
 2\left< K_{\rm p} \right> = -\left< V_{\rm p} \right>,
\label{viri-rela}
\eea
where angle brackets represent the long-time average.
\footnote{
Outside of the above region in SGR model, 
this form of virial relation would be modified 
even in the limit of $\xi \rightarrow 0$,as is shown in Appendix A.}
According to this virial relation, 
the typical size of a system $r$ is related with the total energy $E$ as 
\be
E = {V_{\rm p} \over 2} 
  \equiv -{Gm^2N^2 \over 4r}.
\ee
Hence the above condition for $r$ (\ref{xi<<r<<R}) reads 
\be
\frac{Gm^2 N^2}{4 \xi} \gg  |E| \gg {Gm^2N^2 \over 4R }.
\label{virconE2}
\ee
Moreover from the virial relation, 
the velocity dispersion is given as
\be
2\left< K_{\rm p} \right>   = 
  -\left< V_{\rm p} \right> =
  mN\left< v^2 \right>,
\ee
where $v\equiv dr/dt$ and therefore 
\be
\sqrt{\left< v^2 \right>} = \sqrt{{2|E| \over mN}}. 
\label{vpass}
\ee
{}From this equation (\ref{vpass}), 
the crossing time is defined as 
\be
 \tauc 
  = {r\over\sqrt{\left< v^2 \right>}} 
  = {Gm^{5/2}N^{5/2} \over 4\sqrt{2}|E|^{3/2}}.
\ee
This is the intrinsic time scale associated with genuine gravity 
independent of the cutoff $\xi$ and of the system size $R$.  
\footnote{
As we will show in Sec.\ref{sec:motion}, 
this is also the time scale for the particles 
in \halo species in SGR model.}

The above introduced cutoff parameter $\xi$ connects 
the two limiting cases in the following sense. 
In the limit $\xi \rightarrow 0$, 
the SGR model becomes genuine 3-D gravity at small scales,  
while in the limit $\xi \rightarrow R$, 
the model almost becomes the HMF model. 
The latter is because, in this limit, 
the shape of the potential around $\theta=k\pi (k=0,\pm 1,\cdots)$ 
becomes almost identical as does in the HMF model
\footnote{
The exact HMF model is reproduced in the limit $\xi \rightarrow \infty$.  
}.  

For numerical simulations, 
we need to make all physical variables non-dimensional; 
we use $m$, $R$, and $\tauu$ 
for the unit of mass, distance, and time, respectively.
In these units, 
the physical Hamiltonian (\ref{canoeq}) reads 
\be
 H_{\rm p}= Gm^2{N\over R}H,
\label{H-H}
\ee
where  
\be
H = \frac{1}{2}\sum_{i=1}^N p_i^2
    -\sum_{i<j}^N\frac{1}{\sqrt{2}N\sqrt{1-\cos\theta_{ij}+\epsilon}}.
\label{canoeq-non}
\ee
The dimensionless momentum $p_i$ is given by $p_i = d \theta_i/d\tau$ 
and the dimensionless time $\tau$ is introduced as 
$\tau\equiv t/\tauu$. 
The above form of Hamiltonian 
permits us to use a
powerful symplectic integrator \cite{symplectic}, 
with which the total energy is conserved with extremely high accuracy, 
even beyond thousands of dynamical time.
Typical magnitudes of errors for 
the total Hamiltonian $H(\tau)$ and the total momentum $P(\tau)$ 
in our simulations up to
$\tau \sim 10^{4}$, are $(H(\tau)-H(0))/H(0)\sim {\cal O}(10^{-5})$ 
and $P(\tau)/P_{\mbox{\it r.m.s}}(\tau)
\sim {\cal O}(10^{-8})$, 
respectively, where 
$P_{\mbox{\it r.m.s}} (\tau)\equiv \left(\sum_{i=1}^N p_i^2\right)^{1/2}$.

\section{Classification of phases and particles}
\label{sec:class}
We now study the quasi-equilibrium state 
that appears in a transient stationary stage in our SGR system. 
Though this state is not absolutely stable, 
it generally appears in SGSs during sufficiently long time 
before the system finally approaches the equilibrium state 
characterized by the equipartition of particle energy.

We would like to extract universal properties observed 
in this transient state; 
only the transient description is possible and necessary 
to explain observations.

\subsection{Negative specific heat of SGS}

\begin{figure}[htbp] 
\begin{center} 
 \leavevmode
 \epsfysize=5.0cm
 \epsfbox{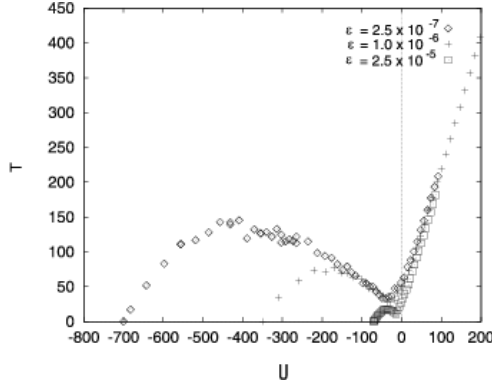}
 \caption{
Diagram of temperature $T$ vs. energy per particle $U$ 
for three different cutoffs 
$\epsilon=2.5\times 10^{-5}$, 
$1.0\times 10^{-6},$ and 
$2.5\times 10^{-7}$. 
In each $T$-$U$ curve, 
there are two critical energy scales; 
$\Ucl$ and $\Uch$, 
between which negative specific heat 
$\partial U/\partial T<0$ appears. 
}
 \label{fig-tu} 
\end{center} 
\end{figure}

First, 
we study the phase diagram of SGR model 
in the temperature-energy plane. 
The $T$-$U$ relation is shown in Fig.\ref{fig-tu} 
and we observe that the region with negative specific heat, 
i.e. negative slope region, apparently exists. 
The temperature $T$ and the internal energy 
per particle $U$ of the present system are defined respectively by 
\bea 
T &\equiv& {2\left<K(\tau)\right>\over N} ~~, 
\label{tempdef}\\
U &\equiv& {H\over N} ~~. 
\eea

\begin{figure}[htbp] \begin{center} \leavevmode
\epsfysize=5.0cm
\epsfbox{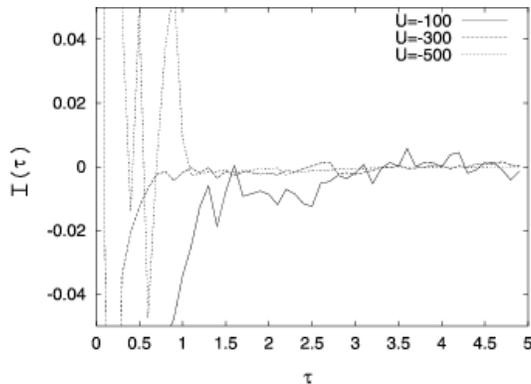}
\caption{
The time evolution of $I(\tau)$ defined by the left side of 
(\protect{\ref{vircon}}) in the case with 
$U=-100$, $-300$, $-500$ and $\epsilon=2.5\times 10^{-7}$.
In each case, 
the system achieves virial equilibrium within a few dynamical time $\tau$.
}
\label{fig-virial} \end{center} \end{figure}

The internal energy $U$ is conserved 
in the present microcanonical system, 
while the temperature $T$ is defined 
as the twice of the time-averaged (represented by angle brackets) 
kinetic energy per particle.
The phase diagram in Fig.\ref{fig-tu} is time averaged until $\tau=2$ 
where the system archives virial equilibrium ({\it see} Fig.\ref{fig-virial}).
Throughout this paper, except explicitly mentioned, 
we fix the total number of particles $N=100$ for simplicity 
and therefore the remaining relevant parameters
which characterize the quasi-equilibrium state of the system would be
$\epsilon$ and $U$.

{}From Fig.\ref{fig-tu}, 
we find two characteristic energy scales 
where $\partial T/\partial U=0$; 
$\Ucl$ at a low energy side and 
$\Uch$ at a high energy side.   
The energy scale $\Uch$ corresponds 
to a mean gravitational binding energy per particle, 
which is estimated as 
$\left< 1/(\sqrt{2}N\sqrt{1-\cos\theta +\epsilon})\right> 
\times (N(N-1)/2) /N \sim {\cal O}(1)$,
while $\Ucl$ strongly depends on the cutoff $\epsilon$. 
This cutoff dependence can be estimated 
from the condition (\ref{virconE2}) 
which, in our normalization, becomes
\be
 -\frac{1}{4\sqrt{2 \epsilon}} \ll U \ll -{1\over 4}.
\label{rangeel}
\ee
Under this condition, 
the negative specific heat condition in virialized state 
would be justified. 
Actually, 
substitution of three different cutoffs 
$\epsilon=2.5 \times 10^{-7}$, $1.0 \times 10^{-6}$, and 
$2.5 \times 10^{-5}$ into (\ref{rangeel}) yields 
the lower limits of (\ref{rangeel}) $353$, $177$, and $35.3$, respectively.
Thus the condition (\ref{rangeel}) correctly describes 
the region of negative specific heat 
in Fig.\ref{fig-tu} with sufficient accuracy. 
However, 
the slope of each T-U line in Fig.\ref{fig-tu} 
is less steep than the value $-2$ 
which is expected for the virial condition of 3-D gravity. 
This discrepancy is probably 
because the energy range satisfying the condition (\ref{rangeel}) 
is too narrow for the ideal T-U curves 
according to our choice of the cutoff parameters; 
even the smallest cutoff we took may not be enough 
to make the relation converge.  

In the range between these two energy scales, i.e. $\Ucl <U< \Uch$, 
the system has  negative specific heat, 
which suggests the existence of phase transition
\cite{Aronson72,Thirring71,Compagner89,Aizawa00}. 
Actually in the system with  negative specific heat, 
a slight energy injection from outside decreases 
the system's temperature and 
induces further energy flow from outside. 
Then this catastrophic temperature reduction induces 
rapid cluster formations in the system. 
As we will see soon below, 
such phase transition from the gaseous state to the cluster state 
actually appears and characteristic structures are realized 
in this intermediate energy range.   

\subsection{Three phases in SGR model }

As is seen in Fig.\ref{fig-tu}, 
there are apparently three phases according to 
the energy per particle $U$; 
a) low energy collapsed phase (\C-phase) $U<\Ucl$, 
b) intermediate energy phase (\I-phase) $\Ucl<U<\Uch$, and 
c) high energy gaseous phase (\G-phase) $\Uch<U$.
The \G-phase (c) is stabilized 
by the infrared cutoff 
($\theta \leq 2\pi$, or a largest physical scale $\sim R$), 
without which the particles would escape into spatial infinity. 
The \C-phase (a) is stabilized 
by the ultraviolet cutoff $\epsilon$, 
without which the particles would fall into a singularity. 
The specific heat for these particles (a and c) is positive, 
in accordance with the stability of these phases. 
On the other hand, the \I-phase (b) (the most specific to gravity) has 
negative specific heat and therefore is unstable. 
The nature of this phase is independent of any artificial cutoffs 
and therefore is thought to represent intrinsic properties of gravity.  

The existence of three different phases, as is explained above, 
distinguishes the SGR model from other models with long-range force. 
For example, 
HMF model has only high and low energy phases, 
and shows a second-order phase transition between these phases. 
On the other hand in SGR model, 
the intermediate phase with negative specific heat exists 
and is strongly unstable when the system is in contact with a heat bath. 
As for 1DS model, 
only a single phase exists 
because the system has no characteristic energy scale; 
there is no phase transition in this model.

The limit $\epsilon \rightarrow 0$ in SGR model represents 
genuine gravity without collision 
and the range of the intermediate phase increases without bound.

\subsection{Three species: \gas, \halo, and \core particles}

In each phase, 
the particles of the system prevail in various energy ranges. 
For example, 
in the intermediate phase, 
some particles evaporate from a cluster 
and move along the ring almost freely with the time scale $\tauu$, 
while some of the others fall into the center of a cluster 
and oscillate with the time scale $\tauf$. 
Thus, the overall phase information does not precisely specify 
the nature of individual particles. 
In order to obtain much fine information, 
we define three species in each phase 
by using the energy of the particles: 
The energy of the $i$'th particle is given by
\be
E_i \equiv 
    \frac{1}{2} p_i^2
    -\sum_{j \neq i}^N\frac{1}
     {\sqrt{2}N\sqrt{1-\cos\theta_{ij}+\epsilon}}.
\label{totali}
\ee
The classification is based on particle energy, that is,
a) \core particles for $E_i<\Ucl$, 
b) \halo particles for $\Ucl<E_i<\Uch$, and 
c) \gas particles for $\Uch<E_i$.
\footnote{
Note that even in the \C-phase, 
there exist a few \gas particles. 
We term the \core particles and \halo particles 
as a cluster in this paper; 
they form an apparent single bound state.
}

\begin{figure}[htbp] \begin{center} \leavevmode
\epsfysize=15.0cm
\epsfbox{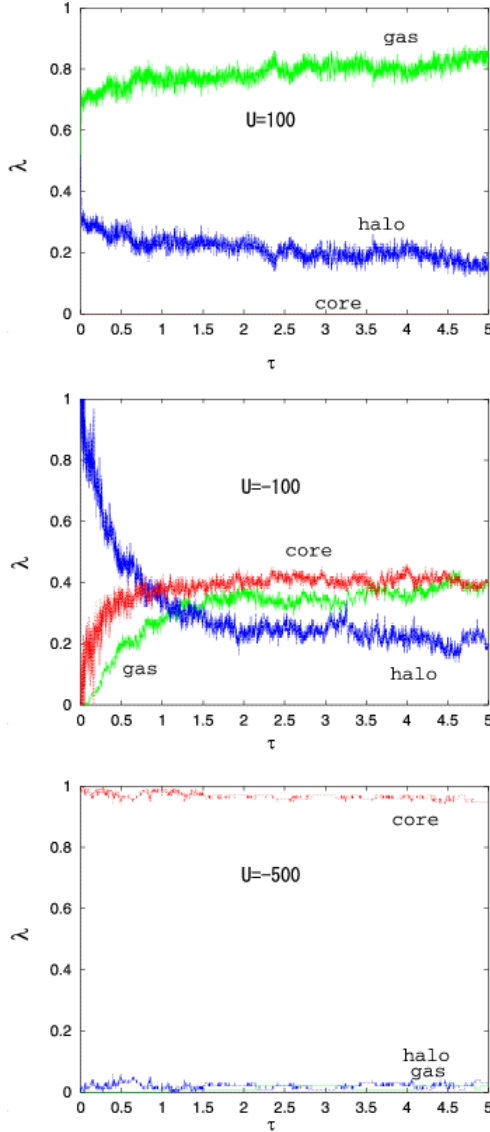}
\caption{
The time evolution of the percentage of each species; 
$\lc$, $\lh$, and $\lg$ 
in high-energy (\G-) phase (top), 
in intermediate (\I-) phase (middle), and 
in low-energy (\C-) phase (bottom) 
for $\epsilon = 2.5\times 10^{-7}$. 
{\it Gas} particles dominate in the high-energy phase, and 
\core particles dominate in the low-energy phase. 
On the other hand, 
all three species coexist in the intermediate phase.}
\label{G-state-t} \end{center} \end{figure}

In the low-temperature phase of SGR, 
the quasi-equilibrium state at very low temperature 
is highly inhomogeneous 
and a single cluster is formed. 
Most particles are condensed 
in this cluster and the total potential is deep. 
In the intermediate phase, 
many particles spread around a cluster. 
As the temperature increases within this phase, 
\halo particles gradually dominate \core particles 
and eventually there appear \gas particles, 
which evaporate from a cluster and go along the ring. 
In the high-temperature phase, 
all particles move almost freely without forming 
any cluster.

We describe the ratio of particle number 
in each states
\footnote{
We term a state to represent the situation
that a particle belongs to one of the three species.
}
by $\lc,$ $\lh,$ and $\lg$, 
with $\lc+\lh+\lg=1$. 
Their evolution is shown in Fig.\ref{G-state-t} 
for the case $\epsilon = 2.5\times 10^{-7}$. 
{}From this, 
we observe those ratios seem to have approached asymptotic values 
beyond $\tau \approx 1$: 
$\lc \approx 0$, $\lh\approx 0.2$, and $\lg\approx 0.8$ 
for $U=100$ (\G-phase), 
$\lc\approx 0.41$, $\lh\approx 0.23$, and $\lg\approx 0.36$ 
for $U=-100$ (\I-phase), 
$\lc\approx 0.96$, $\lh\approx 0.02$, and $\lg\approx 0.02$
for $U=-500 $ (\C-phase). 
{\it Gas} particles dominate in \G-phase and  
\core particles dominate in \C-phase in number.  
While in \I-phase, 
all of the three species of particles coexist almost equally.
According to our various calculations changing the particle number and initial conditions, 
this coexistence seems to represent the prominent property of SGS
and not the finiteness of particles.

\begin{figure}[htbp] \begin{center} \leavevmode
\epsfysize=15.0cm
\epsfbox{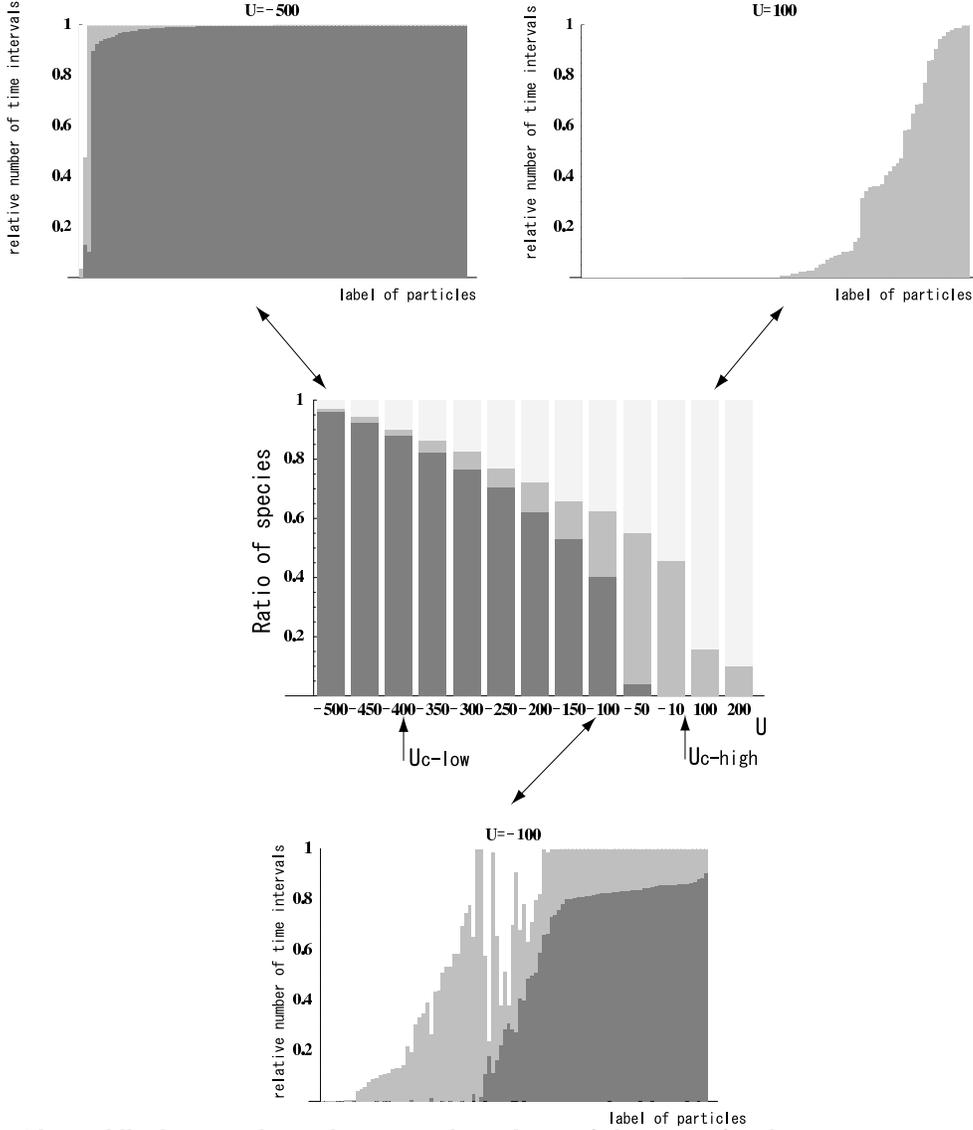}
\caption{
The middle diagram shows the energy dependence of the ratio of each states 
at time $\tau=5$ for the cutoff $\epsilon=2.5\times 10^{-7}$.
For three typical cases, $U=100$, $-100$, and $-500$, 
the ratio of time duration intervals of the three states
in which each particle stays until $\tau=5$ is shown 
in the upper and lower diagrams.
In these diagrams, each vertical bar corresponds to a particle, 
and each ratio of \gas, \halo, and \core state is 
represented by white, gray, and black area, respectively. 
In both \G-phase ($U=100$) and \C-phase ($U=-500$), 
most particles stay within each state.
On the other hand, in the \I-phase ($U=-100$),
many particles experience at least two states.
Some particles wander in all three states.
}
 \label{state-p} \end{center} \end{figure}
In Fig.\ref{state-p},
the energy dependence of the ratio of each species 
in $\tau=5$ for $\epsilon=2.5\times 10^{-7}$ is shown at the center.
The energy scale where \halo particles exit corresponds to 
the one where  negative specific heat appears (\I-phase).
Together with it,
the relative ratio of time intervals of three states 
in which each particle stays until $\tau=5$ is shown. 
As we expect, in both \G- and \C-phase, 
most particles stay just in one state for quite a long time.
For the \I-phase, however, 
many particles experience at least two states.
Some particles wander from one state to another in three states.

\section{Particle motions}
\label{sec:motion}

In this section, 
we examine individual particle motion and 
velocity distribution function in each phase. 

\subsection{Recurrent motion of \halo particles}
\label{sec:rec-motion}

In the HMF model at the state near the critical energy, 
L\'evy-type flight and anomalous diffusion of particles 
have been reported \cite{Latora99}.  
These peculiar behaviors of particles are apparently caused 
by the transitions from \core particles to \gas particles 
and vice versa. 
We observe that such peculiar behaviors in HMF model are caused 
by the periodicity of the configuration space and 
not by the long range force itself. 
In fact, 
for all L\'evy-type flights numerically shown in HMF, 
the flight distances turn out to be longer than the period $2\pi$. 
Therefore the artificial periodicity in the potential, 
and not the long-rage nature of the force itself, 
is thought to have caused the L\'evy-type flight in HMF model. 
In our SGR model, 
we do have the possibility to observe 
the same L\'evy-type flight motions 
since particles in SGR also move along a closed ring. 
Since we would like to extract 
the intrinsic property of the long-range nature of the gravity itself, 
we pay attention to the recurrent motion of \halo particles 
and disregard the round-trip motion along the entire ring.  

\begin{figure}[htbp] 
\begin{center} 
 \leavevmode
 \epsfysize=15.0cm
 \epsfbox{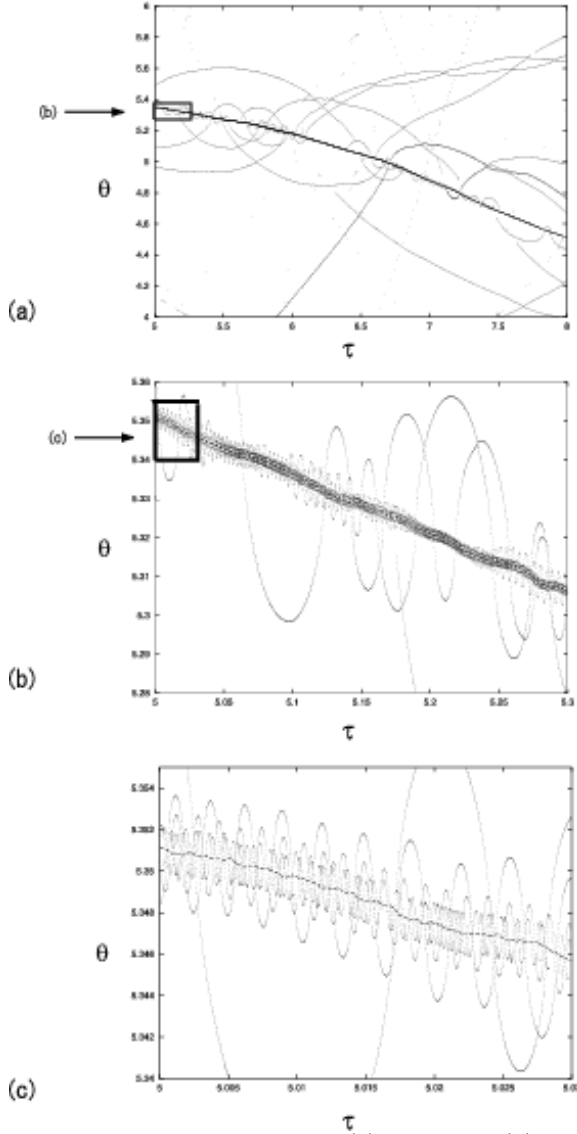}
 \caption{
Trajectories of \halo particles for the time interval 
(a) $\tau=5 \sim 8$, 
(b) $\tau=5 \sim 5.3$ and 
(c) $\tau=5 \sim 5.03$. 
Ten-times zooming up the marked square region in (a) yields (b). 
Further ten-times zooming up the marked square region in (b) 
yields (c). 
The recursion profiles are similar despite the scale difference.}
 \label{fig-stime} 
\end{center} 
\end{figure}

First, 
we choose the parameters as $\xi \approx 7.1 \times 10^{-4} R$ and $U=-100$ 
for which the system is in the intermediate energy phase. 
The particle motion is shown in Fig.\ref{fig-stime}(a), 
in which \core particles form a firm cluster and they oscillate 
around the center of the cluster with the time scale $\tauf$ and 
\halo particles go in and out of the \core region 
without any typical time scale and amplitude. 
Zoomed in ten-times (Fig.\ref{fig-stime}(b)), 
and even in hundred-times (Fig.\ref{fig-stime}(c)), 
the particle motion is always similar recurrent movements. 
These repetition of the similar recurrent pattern suggests 
a self-similar structure of the system. 

Note that the recurrent motion of \halo particles is confined 
within the range $[0,2\pi]$ and 
particles never experience a round-trip along the ring, 
quite contrary to the HMF model. 
Moreover, 
this recurrent motion of particles is quite robust 
and is observed in any region of the intermediate energy phase. 
This robustness is a remarkable contrast to the HMF model 
in which such motion is observed 
only at the critical point in the phase diagram.
\footnote{
The smaller the cutoff $\epsilon$, 
the larger the range of the total energy per particle $U$ 
where the recurrent motion appears.}

In order to analyze this behavior more quantitatively, 
we examine the frequency distribution of the recurrent time 
$\tau_{\rm rec} \equiv \tau_{\rm in}-\tau_{\rm out}$, 
that is, the time period 
from the moment $\tau_{\rm out}$ 
when a particle leaves the barycenter of the \core 
to the moment $\tau_{\rm in}$ 
when it first returns to the barycenter again. 
Here we have defined the location of the barycenter of the \core as 
\be
\theta_{bc} \equiv
\sum_{i=1}^{N_c}\theta_{i}/N_c ,
\ee
where $\theta_{i}$ is the location of $i$'th particle in the \core 
and $N_c$ is the total number of \core particles at each moment.

\begin{figure}[htbp] 
\begin{center} 
 \leavevmode
 \epsfysize=5.0cm
 \epsfbox{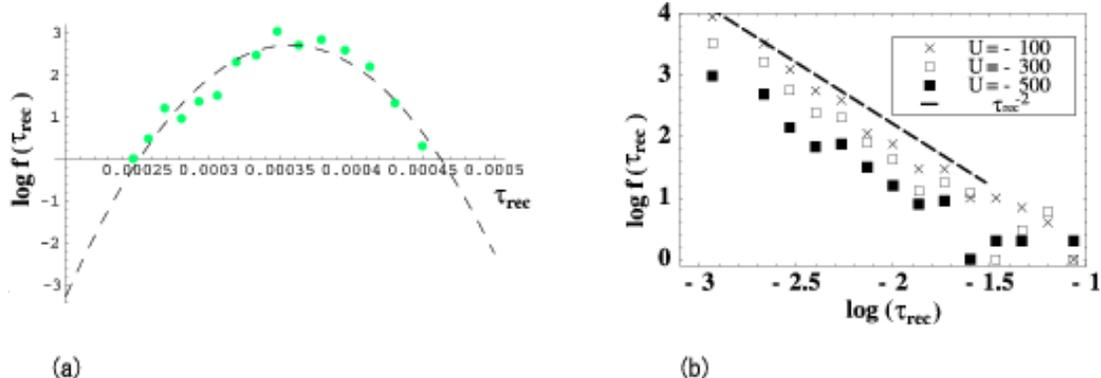}
 \caption{ 
A linear-log plot of frequency distribution 
$f(\taurec)$ for \core particles and 
a logarithmic plot of frequency distribution 
$f(\taurec)$ for \halo particles in several values of $U$. 
Each plot in (b) shows the same slope 
within the range $10^{-3.0} <\taurec< 10^{-1.5}$. 
The broken line represents a line with slope $-2.0$.}
 \label{fig-freq-ch} 
\end{center} 
\end{figure}

In Fig.\ref{fig-freq-ch}, 
we depicted the frequency distribution of the recurrent time 
$\taurec$ for \core particles and \halo particles separately. 
For \core particles, as we expected, 
we find almost Gaussian distribution around the center $\tautf$
(Fig.\ref{fig-freq-ch}(a)).
On the other hand for \halo particles, 
although a peak is found around $\tautc (\equiv \tauc/\tauu)$, 
the distribution shows that a long tail spreads widely characterized 
by the power law $f(\taurec)\sim \taurec^{~-p}$ with $p \approx 2.0$ 
(Fig.\ref{fig-freq-ch}(b)). 
We have checked 
that this power law range $-3<\log\taurec<-1.5$ is consistent 
with the eye-fitted region of self-similar motion in Fig.\ref{fig-stime}.
We will see that the above difference of frequency distribution 
for \core and \halo particles leads to the difference 
in relaxation time of them in Sec.\ref{sec:relax}.

We have also examined the energy dependence of the above power $p$. 
The result is given in Fig.\ref{fig-freq-ch}(b) 
showing that the value of $p$ ($\approx 2.0$) is almost 
independent of the choice of the energy $U$ 
throughout the intermediate energy phase.

\subsection{Velocity distribution of particles}
\label{sec:veldist}

As is shown in Appendix \ref{sec:viricon}, 
the time averaged kinetic energy $\left<K\right>$ is, in general, 
expressed as the sum of $N^2$ independent stochastic variables. 
However, 
in the case of \core particles, 
which are located within a few cutoff scale, 
the quantity $\left<K\right>$ turns out to be expressed 
as the sum of $N$ independent stochastic variables 
({\it see} Appendix \ref{sec:viricon-core}). 
In this section, 
we show that the $N$-dependence  of $\left<K\right>$ characterizes 
the velocity distribution of particles.  

\begin{figure}[htbp] \begin{center} \leavevmode
\epsfysize=10.0cm
\epsfbox{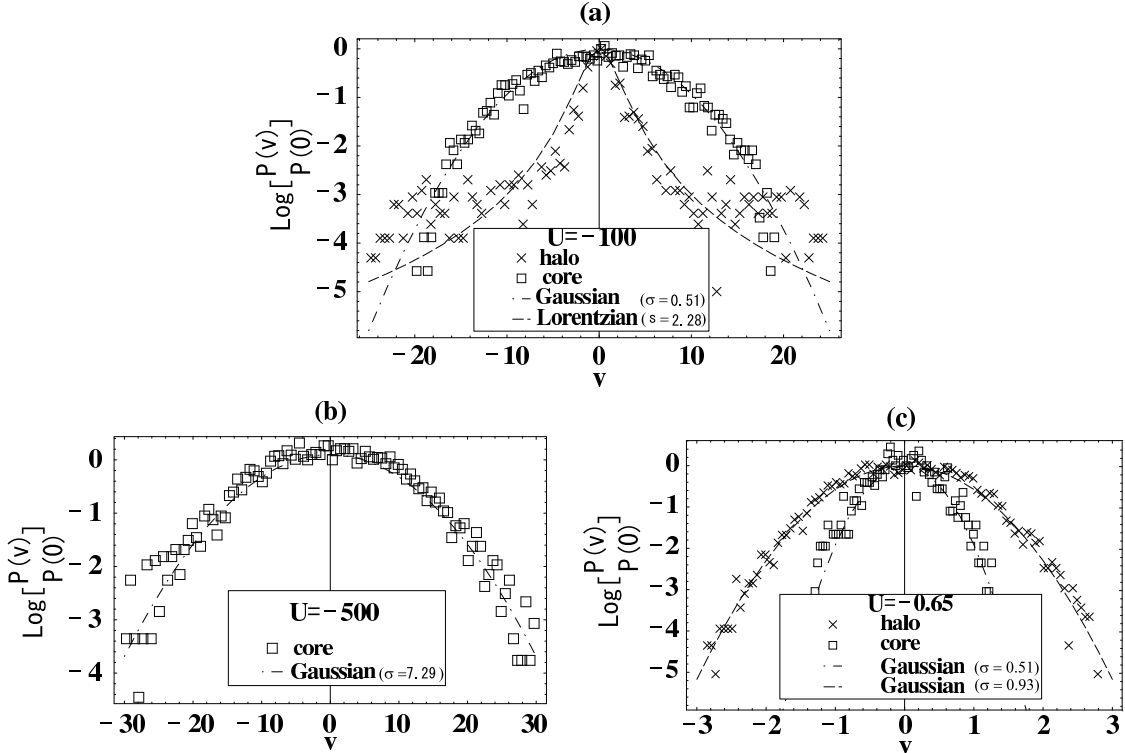}
\caption{
The linear-log plot of velocity distributions of particles 
for the three cases; 
(a)$\epsilon=2.5\times 10^{-7}$, $U=-100$ (\I-phase), 
(b)$\epsilon=2.5\times 10^{-7}$, $U=-500$ (\C-phase), and
(c)$\epsilon=2.5\times 10^{-3}$, $U=-0.65$ (\I-phase).  
In all cases, 
velocities of \core particles are well fitted 
by the Gaussian distribution with dispersion 
$\sigma=0.51$(a), $7.29$(b), and $0.51$(c) respectively. 
Also in the large cutoff case (c), 
velocities of \halo particles are well fitted 
by the Gaussian distribution $\sigma=0.93$.
On the other hand, 
in the small cutoff case (a),
velocities of \halo particles are well fitted 
by the Lorentzian distribution ($\alpha=1$) with $s=2.2$.}
\label{fig-v} \end{center} \end{figure}

In Fig.\ref{fig-v}, 
we show the velocity distributions 
of \core and \halo particles in three cases; 
(a) $\xi \approx 5 \times 10^{-4} R$, $U=-100$ 
(b) $\xi \approx 5 \times 10^{-4} R$, $U=-500$ and 
(c) $\xi \approx 5 \times 10^{-2} R$, $U=-0.65$. 
We have superposed the velocity data at $\tau=1$, $2$, $3$, $4$, and $5$, 
with ten different random initial conditions fixing the total energy. 
Thus the size of the whole data we used is $5 \times 10^3$ 
for each velocity distribution function. 
The distribution of \core particles is well fitted 
by the Gaussian distribution
\be
P(v) = {1\over\sqrt{2\pi} \sigma} e^{-{v^2 \over 2\sigma^2}},
\label{gaussian}
\ee
with the dispersion 
$\left<v_{\core}^{~2}\right>^{1/2}\equiv 
\sigma=7.29(a), 10.9(b)$, and $0.51(c)$. 

On the other hand, 
the distribution of \halo particles in (a) is manifestly non-Gaussian, 
although that of the same particles in the case (c) is Gaussian.  
This distribution is at least unchanged until $\tau=200$ .

What is the origin of the above non-Gaussian distribution 
for \halo particles in the intermediate energy phase in case (a)? 
Here we analyze this issue 
from the viewpoint of the generalized central-limit theorem. 
It is well known 
that the limit distribution for sums of independent random variables is 
the Gaussian distribution 
provided that the dispersion is finite (the central-limit theorem). 
However, 
it is less well-known that 
the limit distribution for sums of independent random variables is 
the stable distribution in general cases, including those 
where the dispersion is divergent. 
This stable distribution is defined to satisfy the relation 
\be
\sum_{i=1}^{N} x_i \stackrel{d}{=}  N^{1/\alpha} x, 
\ee
where $\stackrel{d}{=}$ means 
that the distributions of both sides are equal to  each other. 
The parameter $\alpha$ classifies the stable distributions 
and must satisfy $0<\alpha \leq 2$ for the normalizability 
and the positivity of the probability distribution function. 
In the definition above, 
$x_i$ and $x$ are the probabilistic variables 
obeying the same distribution. 
Let us apply this to the velocity distribution of our model. 
We square both sides of the above equation, 
and obtain the non-extensive property,
\be 
\left<v^2\right>_{N{\rm -particle}} = 
   N^{2/\alpha-1}\left<v^2\right>_{\rm one-particle}
\label{bunsan}
\ee
and the method of characteristic function yields the explicit 
form of the distribution function \cite{Feller66}
\be
P(v) = 
  {1\over  2\pi} \int^\infty_{-\infty} dy 
   ~\exp[-ivy-s|y|^\alpha ].
\label{stable}
\ee
Note that this stable distribution includes 
the Gaussian  distribution (\ref{gaussian})
as a special case $\alpha=2$, where 
the kinetic energy becomes extensive, and 
the dispersion $\sigma =\sqrt{2s}$ is finite. 
The parameter $s$ is thought to be a generalized temperature.  

Let us first consider the velocity distributions of \core particles. 
We observe, from our numerical calculations, 
that all \core particles oscillate 
within the narrow region of the cutoff size $\xi$, 
and the gravitational two-body interaction is dominated 
by the artificial potential force. 
Therefore the averaged kinetic energy is described 
as the sum of $N$ independent statistical elements 
(Appendix \ref{sec:viricon-core}). 
This leads to the normal extensivity 
for the velocity distributions of \core particles 
and therefore we expect the Gaussian distribution. 
This is consistent with the results in Fig.\ref{fig-v}.   

On the other hand, 
for \halo particles which interact through genuine gravity, 
the kinetic energy behaves 
as $\left<K_{\rm p}\right>\sim N^2$ for fixed $R$ 
as is shown in Appendix \ref{sec:viricon}. 
Thus from (\ref{bunsan}) we find the index $\alpha=1$ 
for the physical velocity distribution of SGR. 
In this case with  $\alpha=1$, 
the distribution (\ref{stable}) becomes the Lorentz form
\be
P(v) = {1\over \pi}{s \over v^2+s^2}.
\label{dis1}
\ee
This is also consistent with the results in Fig.\ref{fig-v}. 
The essence of the appearance of this non-Gaussian distribution is 
the non-extensivity of the energy for SGS; this is 
the intrinsic property of gravity.   

Then what is the origin of the Gaussian distribution 
of \halo particles in case (c)?
As we show in the next section, 
this is mainly because the relaxation times 
of \core and \halo particles are close with each other.
In the next section, we will discuss the relation between the profile
of velocity distribution and the relaxation process. 

\section{Relaxation}
\label{sec:relax}

One of the most important issues 
in statistical physics of the $N$-particle system is the relaxation process.
In order to study the relaxation process of our model, 
we choose several different initial conditions 
for the same values of $\epsilon$ and $U$. 
Here we have examined two parameter sets, i.e. 
(i) $\epsilon = 2.5 \times 10^{-3}, U=-0.65$ and 
(ii) $\epsilon = 2.5 \times 10^{-7},U=-100$, 
for both of which negative specific heat appears. 

As we discussed in Sec.\ref{sec:SGR}, 
there are three dynamical time scales: 
$\tauu$, $\tauf$, and $\tauc$. 
In the units of our normalization, 
the ratios of these time scales are
\bea
\tauc/\tauu&=&{1 \over 4\sqrt{2}}|U|^{-3/2} \non \\
\tauf/\tauu&=&(2 \epsilon)^{3/4}.
\label{tratios}
\eea
For example, if we choose $U=-100$ and $\epsilon = 2.5\times 10^{-7}$, 
then $ \tauc/\tauu \approx 1.8 \times 10^{-4}$ and therefore
the three time scales are separated as
\be
\tauf<\tauc \ll \tauu.
\ee
Then the contribution of \gas particles to the dynamics of \halo
or \core particles is negligible during a few $\tauu$.

The parameter set
(i) does not meet the condition(\ref{rangeel}) 
and $\tauc/\tauu\approx 0.34$; $\tauc$ is the same order as $\tauu$.
On the other hand, the parameter set 
(ii) meets (\ref{rangeel}) and 
$\tauc/\tauu\approx 1.8 \times 10^{-4}$; $\tauc$ is much shorter than $\tauu$.
This means that \gas species play an important
role in relaxation in case (i), while they do not
in case (ii).  In order to examine the relaxation process
for these two cases,
it is convenient to normalize the 
time with the unit $\tauc$, since the degree of 
relaxation of 3-D gravitational system
is conventionally
measured with the dynamical time. 
So here we introduce the dimensionless time
$\tautc \equiv t/ \tauc$.

As for initial conditions, 
we locate several clusters of the same size with the same interval. 
The number of clusters is chosen as
5, 10, 20, 25, and 50 for the model (i)
and 5, 10, 20, and 25 for the model (ii).
The initial velocities of particles are set randomly but the total energy 
is fixed: $U=-0.65$ for (i) and $U=-100$ for (ii).

In order to study the Ergode 
property of the system, we examine the mixing property of the particles 
in the three states( \core, \halo, and \gas).  In other words, we examine the 
degree of isolation of each states.  
We directly measure how extent each particle experiences these three states.  
First, we 
pick up discrete times $(\tautcf, \tautcs, \cdots,
\tautci, \cdots)$ with the equal interval $\Delta \tautc$; at each time, each particle stays at one of the three states.
Then we count how many times the $k$-'th particle stays at each states before the time $\tau_{Ei}$.  
Finally, after normalization, 
we obtain the relative frequencies of three states
$\lck (\tautci),\lhk (\tautci)$ and $\lgk (\tautci)$ 
(with $\lck (\tautci)+\lhk (\tautci)$+$\lgk (\tautci)=1$ )
for the $k$'th particle.  

At the first time $\tautcf$, 
each particle definitely stays at one of the three states; relative frequencies of stay 
$\lck (\tautcf),\lhk (\tautcf)$ and $\lgk (\tautcf)$ are either 1 or 0. 
In the next time $\tautcs$, some particles may
change the state; then $\lck (\tautcs),\lhk (\tautcs)$ and 
$\lgk (\tautcs)$ for these particles are $1/2$. 
In this way, $\lck (\tautci),\lhk (\tautci)$ and $\lgk (\tautci)$ 
will evolve in time $\tau_{Ei}$. 
Various distribution of $\{\lck (\tautci)\}_{k=1}^{N}$ for all the particles 
$k=1,2,\cdots,N$ will define the distribution function $N(\lg,\:\tautc)$ 
for values $\lambda_{gas}$ at time $\tau_{Ei}$ by simply counting 
the number of the particles which take the value $\lc$ at time $\tautc$.  
Similarly distribution functions for other states can also be defined: 
$N(\lh,\:\tautc)$ and $N(\lc,\:\tautc)$.  
As the system is thermally relaxed and 
the equipartition of energy is attained,
$\lck (\tautc),\lhk (\tautc)$ and $\lgk (\tautc)$ will
converge to the same value $\lc^{\ast},\lh^{\ast}$ and
$\lg^{\ast}$ independent of the particle label $k$. 
Then $N(\lg,\:\tautc)$, $N(\lh,\:\tautc)$ and $N(\lc,\:\tautc)$ 
would have sharp peaks at $\lc^{\ast},\lh^{\ast}$ and
$\lg^{\ast}$ in the thermal equilibrium state. 
Thus the development of the peaks 
in the distributions functions $N(\lg,\:\tautc)$, 
$N(\lh,\:\tautc)$ and $N(\lc,\:\tautc)$ can be a good measure 
of the degree of thermal equilibrium. 

\begin{figure}[htbp] \begin{center} \leavevmode
\epsfysize=12.0cm
\epsfbox{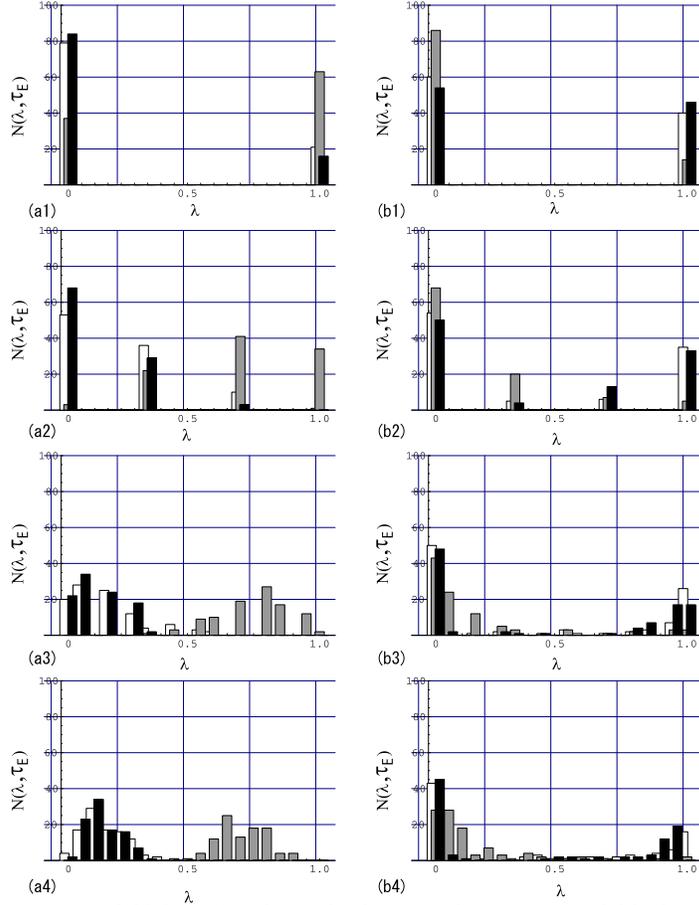}
\caption{
The time variation of $N(\lambda,\:\tautc)$, 
where the bar in white, gray and black colors show
$N(\lg,\:\tautc)$, $N(\lh,\:\tautc)$
and $N(\lc,\:\tautc)$, respectively.
The left and right  columns respectively   represent
the case of large UV-cutoff 
(a) $\epsilon = 2.5 \times 10^{-3}$ and $U=-0.65$ and 
the case of small UV-cutoff 
(b) $\epsilon = 2.5 \times 10^{-7}$ and $U=-100$,
where $\tautcz=6.5\times 10^4$, 
$\Delta \tautc = 2.0 \times 10^3$ for (a), and 
$\tautcz=1.1\times 10^5$, $\Delta \tautc = 3.8 \times 10^3$ for (b)
In both columns, time flows from top to bottom: 
$(a1,b1)\; \tau_{E_1}, 
(a2,b2)\; \tau_{E_3},(a3,b3)\; \tau_{E_{12}},(a4,b4)\; \tau_{E_{34}}$, 
respectively. 
}
\label{equifig} \end{center} \end{figure}

In the large UV-cutoff case (i), 
each distribution $N(\lg,\:\tautc)$, $N(\lh,\:\tautc)$
and $N(\lc,\:\tautc)$ starting from
$\tautcz=6.5 \times 10^4$ 
seems to develop a single peak in-between 0 and 1 at least
by the time $\tautc \approx 10^5$ (Fig.\ref{equifig} (b)),
and the variance around the peak
seems to reduce in time (Fig.\ref{equifig} (c)).
This means that most particles experience all of the three states
at least by $\tautc \approx 2.0 \times 10^5$; thermal relaxation
proceeds in the time scale $\tautc \sim 10^5$.   
Contrary in the small UV-cutoff case (ii), 
each distribution $N(\lg,\:\tautc)$, $N(\lh,\:\tautc)$
and $N(\lc,\:\tautc)$ starting from
$\tautcz=1.1 \times 10^5$
seems to develop peaks at the edge of the domain 1 and/or 0, 
even after the time duration, 
$\tautc \approx 3.8\times 10^5$(Fig.\ref{equifig} (e)) .
This means that most particles stay in each single state even in
the time scale $\tautc \sim 10^5$; the system does not
approach thermal relaxation at all in this time 
scale.

By combining these results,
the Gaussian velocity distribution realized in the case (i) in Fig.\ref{fig-v}
seems to reflect the achieved thermal relaxation of the system.  
On the other hand, 
the Lorentzian velocity distribution realized 
in the case (ii) in Fig.\ref{fig-v}
seems to reflect some quasi-equilibrium intermediate stage 
before the thermal relaxation of the system.  
Latter peculiar properties in velocity distributions 
may make 3-D gravitational system quite 
different from those in the system with short-range interactions or
the system with positive specific heat. 
If we set $\epsilon$ larger, then the encounter of \halo particles 
with normal particles (\gas and \core particles) becomes more frequent, and 
the \halo particles cannot be isolated from the normal particles, resulting 
trivial thermodynamical equilibrium of the whole system as in the case (i). 

\section{Scaling properties of the \HALO}
\label{sec:scaling}

In Sec.\ref{sec:motion}, 
we have found that the velocity distribution of \halo particles is 
non-Gaussian with anomalous $v^{-2}$-tail and 
each \halo particle shows intermittent recursive motion 
around the \core without a definite time scale. 
These scale-free properties of the \halo particles turn out to appear 
in the spatial distribution profile of them itself. 
In this section, 
we study the scaling property of the spatial distribution 
with the box-counting method \cite{box-counting}.

\begin{figure}[htbp] 
\begin{center} 
 \leavevmode
 \epsfysize=5.0cm
 \epsfbox{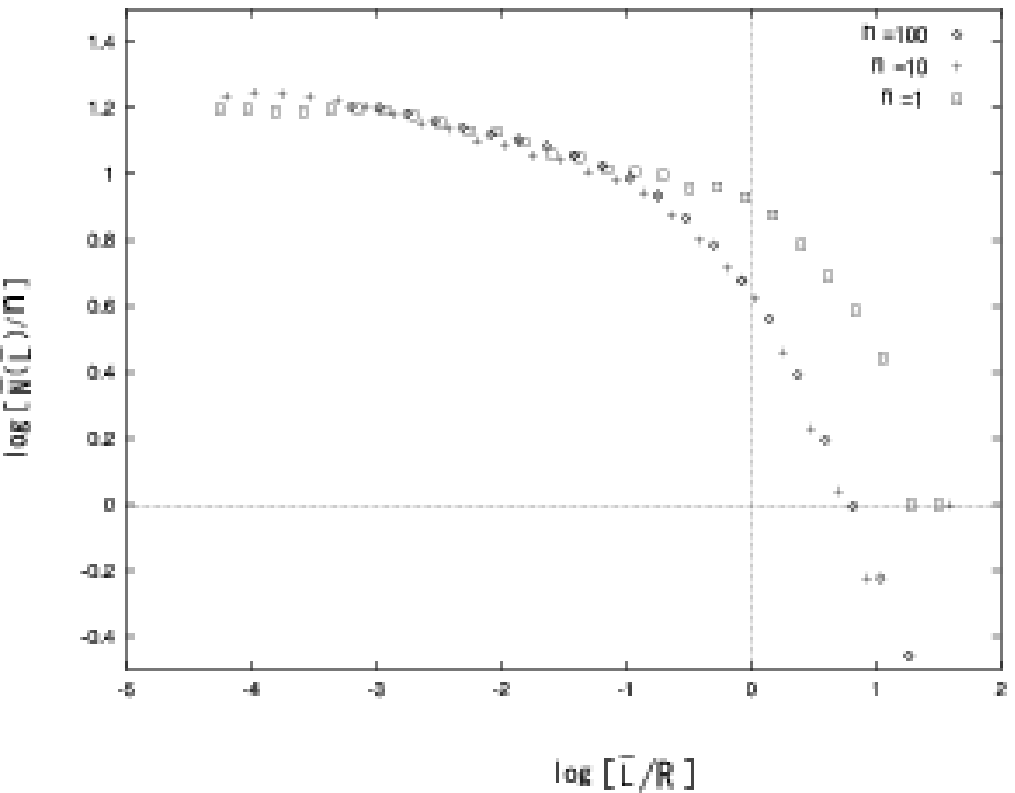}
 \caption{
Log-log plot for 
the box scale $\bar{L}$ vs the number of the box $\bar{N}$ 
occupied by at least one \halo particle. 
We superposed the snapshot data of $n$ different times.
The cases with $n=1$, 
$10~(\tau_0=2.0,\Delta \tau=0.3)$, and 
$100~(\tau_0=2.0,\Delta \tau =0.03)$ 
with the cutoff $\epsilon=2.5\times 10^{-7}$ 
($\xi \approx 7.1\times 10^{-4} R$) are plotted.
The plots of these three cases coincide with each other 
in the range $\xi  < \bar{L} < 10^{-1}R$.
The scaling exponent $D(\bar{L})$ derived from the slope 
in this range is 0.1. 
}
 \label{boxav-loglog} 
\end{center} 
\end{figure}

For this purpose, we use the box counting method: 
We divide the entire configuration space $2 \pi R$ 
into segments with equal size $\ell$ 
and count the number $N(\ell)$ of the segment 
which contains at least one \halo particle. 
Then we define the quantity 
$- \partial (\log N(\ell)) / \partial (\log\ell)$ 
which turns out to be the scaling exponent, 
provided that the quantity is almost independent of the scale $\ell$. 
In order to increase the statistical significance, we used 
multiple particle-distribution data at different times for each run. 
Moreover, we made many runs of calculations 
with different initial conditions for the
fixed total energy $U$. 
We have applied the box-counting method to each data 
and then all the individual results are superposed. 
When we extract the distribution data, 
we have chosen the time interval $[2000\tauf,5000\tauf]$, 
in which the particle ratios $\lc$, $\lh$, and $\lg$ almost 
reach the relaxed constant values. 
Thus we calculate the following averaged quantity
\be
D(\ell) \equiv 
   -\left< { \partial (\log N(\ell)) 
   \over \partial (\log\ell)}\right>_{\rm ens},
\ee
over all the data we thus prepared. 
If any scaling property exists in the particle distribution, 
$D(\ell)$ would become constant 
for a finite range of $\ell_1 \leq \ell \leq \ell_2$. 
As for the bounds $\ell_1$ and $\ell_2$, 
we have technical restrictions originating 
from our numerical calculation method. 
The averaged distance between \gas particles is estimated as
\be
l_{\mbox{gas}}(\epsilon) \sim { 2\pi R \over N\lambda_{\mbox{gas}}}.
\ee
If the box size $\ell$ is larger than $l_{\mbox{gas}}(\epsilon)$, 
we expect $D \approx 1.0$, 
because almost all the boxes contain at least one particle. 
On the other hand, 
if $\ell$ is smaller than the cutoff distance $\xi$, 
the genuine property of gravity is lost. 
Therefore scaling property in particle distribution is 
relevant only within the rage $\xi \leq \ell\leq l_{\mbox{gas}}$. 
Actually in our calculation, 
for $\epsilon=2.5\times 10^{-7}$ and $U=-100$, 
we can see the typical behavior of $N(\ell)$ 
as a function of $\ell$ in Fig.\ref{boxav-loglog}. 
In this plot, 
we can see scaling behavior $N({\ell}) \sim {\ell}^{-d}$ 
with the small exponent in the range $10^{-3}R<\ell < 10^{-1} R$, 
which is well inside the relevant region. 
This scaling seems to originate from \halo particles, 
because they
not only dominate in the above scaling range 
but also show non-Gaussian velocity distributions 
and self-similar recursion jumps.

However, 
the scaling property of the box counting method does not always conclude 
the existence of the fractal structure in the particle distributions. 
This is because the box counting method itself cannot distinguish 
the two possibilities; 
a) genuine fractal structure and
b) the power-law tail of the particle distributions around the \core center. 

In order to distinguish the above possibilities, 
we compare two different superposition methods as follows:
a) We simply superpose all the data with the bare coordinate $\theta$ and 
b) we superpose all the data with the coordinate adjusting 
so that the mass-center of each data comes to the same position; 
i.e. we introduce a new coordinate 
$\bar{\theta}$ as $\bar{\theta} \equiv \theta - \theta_{bc}$ 
for each data with the mass center $\theta_{bc}$.  

As a result, 
the scaling property of the particle distributions, 
which was observed in the original method, 
disappears completely in the case a). 
On the other hand in the case b) 
we observe the same scaling property 
as was observed in the original method. 
This result suggests that 
the observed scaling property in particle distribution is 
not the genuine fractal property of the system, 
but simply due to the power-law distribution of particles 
around the mass-center. 
We have confirmed this with our further analysis 
on the direct observation of the power-law particle distributions 
around the mass-center.  

\begin{figure}[htbp] 
\begin{center} 
 \leavevmode
 \epsfysize=5.0cm
 \epsfbox{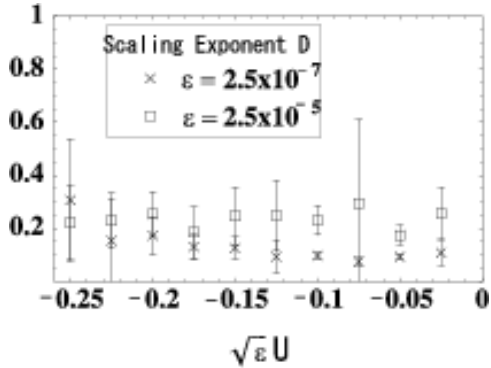}
 \caption{
The scaling exponent D vs the energy per particle $U$.
Most of the energy region shown in the figure corresponds to \I-phase.
For convenience, 
we use $\protect{\sqrt{\epsilon}U}$ as a horizontal axis.
The case of $\epsilon=2.5\times 10^{-5}$ 
(squares with error bars ) and 
the case of $\epsilon=2.5\times 10^{-7}$ 
(crosses with error bars ) are plotted.
It seems that the dimension $D$ does not depend on $U$.
Since the number of \halo particles is small in low energy,
the error becomes large there.}
 \label{fig-frac} 
\end{center} 
\end{figure}

The energy and cutoff dependence on the scaling exponent $D$ 
is shown in Fig.\ref{fig-frac}. 
Here we show the scaling exponent of \halo particles 
in the region of negative specific heat. 
We find that the scaling exponent $D$ is almost constant 
in this region. 
Moreover, 
smaller cutoff $\epsilon$ corresponds to 
the smaller dimension $D$. 
On the other hand in the limit of $\epsilon \rightarrow 1$, 
the scaling property cannot be observed.

In the limit of real gravitational interaction 
($\epsilon\rightarrow 0$), 
the region of negative specific heat extends in energy range 
and so does the scaling region. 
This constant property of the exponent $D$ is likely to be independent of 
the temperature or the energy of the system. 

\section{Conclusion}
\label{sec:con}
In this paper 
we have studied the self-gravitating ring (SGR) model 
which is a one-dimensional (1-D) model 
of a self-gravitating system. 
Although the system is 1-D, 
the particles constrained on a circular ring 
mutually interact with the 3-D full gravitational force. 
We found that an interesting phase with the negative specific heat 
appears at the intermediate energy scale, 
which reflects the virial condition of the 3-D gravitating system. 
Classifying the particles in each phase 
into three species (\core, \halo, and \gas specie), 
we examined dynamical properties of this phase.  
Using SGR model, 
we further examined characteristic properties of relaxation, 
velocity distributions, and density profiles in each phase.

The cutoff parameter $\xi$ introduced in SGR model controls 
the singular properties of the system at short-distances. 
By changing this parameter, 
we could systematically study 
the effect of the short-distance singularity of the force 
upon the statistical properties of the system.   

When the cutoff parameter is of order one ($\xi \sim R$ ), 
the system resembles the Hamilton Mean Field model (HMF). 
In fact,  
the velocity distribution becomes Gaussian 
at $t\sim{\cal O}(10^3)~\tauu$.   
This is because the two time scales $\tauc$ and $\tauu$ become the same order 
and therefore \halo particles frequently encounter and interact 
with \gas particles. 
This strong interaction induces frequent exchange of particles 
between these two species. 
Thus the energy diffuses effectively 
and complete thermodynamical equilibrium of the entire system is established. 
Especially the temperature is the same for all the particle species.  
In other words, 
the negative specific heat region (\halo particles) does exist,  
but the strong interaction with \gas and \core particles 
(positive specific heat) 
quickly dissolves the negative specific heat region. 

In much more smaller cutoff case with 
$\xi \approx 7.1 \times 10^{-2} R (\epsilon = 2.5 \times 10^{-3})$, 
the velocity distribution becomes almost Gaussian 
after the virialization, 
although the temperature of each species is different from each other. 
This temperature difference, 
as well as the required time for establishing the equilibrium, 
is increased for smaller value of the cutoff $\xi$.   

On the other hand, as $\xi \rightarrow 0$, 
the difference of the two time scales increases $\tauu \gg \tauf$, 
and the energy region of the negative specific heat extends wider. 
Therefore we can expect 
that \halo particles become almost isolated from the \gas particles. 
This isolation of unstable \halo particles from other normal species enables 
the \halo species to last sufficiently long.  
In the case of
$\xi \approx 7.1 \times 10^{-4} R (\epsilon = 2.5 \times 10^{-7})$, 
the initial condition dependence of \halo particles survives 
even after $ 5 \times 10^{4}\tauu $ in our simulation, 
which prevents the full system 
from reaching thermal equilibrium. 
These results resemble the case of \core particles 
with $\xi \approx 7.1 \times 10^{-2} R$.
However, 
the remarkable difference between the two cases is that
the velocity distribution of \halo particles 
in $\xi \approx 7.1 \times 10^{-4} R$
is not Gaussian but Lorentzian distribution, 
while that of \core particles in $\xi \approx 7.1 \times 10^{-4} R$
is Gaussian. 
This property of \halo particles qualitatively reflects 
genuine 3-D gravitational interaction
and is the main difference from the HMF models or 1DS models 
We have also found 
the scaling structure of \halo particles 
in the intermediate energy phase. 
When we reduce the cutoff $\xi \rightarrow 0$, 
the scaling region as well as the energy range of negative specific heat 
is extended wider, 
although the value of exponent depends on the cutoff 
and reduces with $\xi$.

In 1DS model, 
the system reaches thermodynamical equilibrium  at least at 
$t\sim{\cal O}(10^7) \tauc$, 
where $\tauc  \equiv (1/4\pi GmN)\sqrt{4E/mN}$ 
is the crossing time of 1DS model\cite{Tsuchiya96}.
It is also shown that 
the system reaches the collisionless mixing phase 
during much shorter time interval. 
Our simulation with $\xi \approx 7.1 \times 10^{-4} R$
shows non-Gaussian velocity and scaling property 
in such a short time interval 
$4 \times 10^3  ~\tauc \sim 2 \times 10^5 ~\tauc$ for $U=-100$.
So the non-Gaussian velocity distribution might reflect
the character in collisionless mixing phase
which appears in 1DS simulation \cite{Tsuchiya96}. 
However, 
it might also be true that the relaxation process in SGR 
is quite different from that in 1DS, 
where no time scale separation between \core and \halo particles exists.
So the exotic character of \halo particles might be
intrinsic in SGR or the system with real 3-D gravitational interactions
because of the time scale separation.

{}
In summary,
SGR model is characterized (i) by  the phase separation and
the particle state separation and (ii) by  
similar recurrent motion and non-Gaussian velocity distribution
of \halo particles . Both of these characters may play a key role 
in explaining 
the observational properties
in real 3-D self-gravitating objects.
For example, 
the recent observations with Hubble Space Telescope support 
the existence of supermassive black holes 
or cusps at the center of  triaxial  elliptical galaxies
\cite{Lauer95,Byun96,Gebhardt96}.
This observation  suggests that
the above property (i)  may
appear in these galaxies, since they have high density region
corresponding to \core particles at their centers and low density
region corresponding to \halo particles surrounding  them. 
Hence, the  property (i)  may  give a hint for explaining
the relaxation process and the origin of stationally configuration
shape of 
these galaxies, since
the stochastic \halo
 orbits which  interact with
\core particles at the center of a triaxial galaxy
 may affect the equilibrium shape 
through the continued  mixing near the center
\cite{Merritt96}.
In addition, 
the propety (ii)  might help  to explain 
the observed fractal structures
and non-Gaussian velocity distributions 
in the inter stellar medium\cite{Falgarone91}, 
which are likely to be gravitationally virialized.

\acknowledgments

We would like to thank A. Nakamichi, I. Joichi, K. Nakamura, 
and M. Hotta for useful discussions and comments. 
This work was supported partially by a Grant-in-Aid 
for Scientific Research Fund of the Ministry of Education, 
Science and Culture (Specially Promoted Research No. 08102010), 
and by the Waseda University Grant for Special Research Projects.

\appendix

\section{Virial condition for SGR model}
\label{sec:viricon}

We start with the non-dimensional Hamiltonian of the SGR model,
\be
H=\frac{1}{2}\sum_{i=1}^N p_i^2
  -\sum_{i<j}^N
  \frac{1}{\sqrt{2}N\sqrt{1-\cos\theta_{ij}+\epsilon}}.
\label{canoeq2}
\ee
Canonical equations of motion for this system become
\bea
\frac{d\theta_i}{d\tau}&=&p_i, \non \\
\frac{dp_i}{d\tau}&=&
 -\sum_{j\neq i}^N\frac{\sin\theta_{ij}}
 {2\sqrt{2}N(1-\cos\theta_{ij}+\epsilon)^{3/2}}.
\label{canoeq3}
\eea
Then the second time derivative of the inertia moment of this system becomes
\bea
 {d^2 \over d\tau^2} \sum_{i=1}^N\theta_i^2 
 &=& 2\left(\sum_{i=1}^N p_i \frac{d\theta_i}{d\tau}+
     \sum_{i=1}^N \theta_i \frac{dp_i}{d\tau}\right)\non \\
 &=& 2\sum_{i=1}^N p_i^2-2\sum_{i<j}^N  \frac{\ \theta_i \sin\theta_{ij}}
     {\sqrt{2}N(1-\cos\theta_{ij}+\epsilon)^{3/2}}.
\eea
Taking long-time average for this equation, we obtain 
\be
\left< {d^2 \over d\tau^2} \sum_{i=1}^N\theta_i^2 \right> 
  = 4NT-2\left< \sum_{i<j}^N  \frac{\ \theta_i \sin\theta_{ij}}
    {\sqrt{2}N(1-\cos\theta_{ij}+\epsilon)^{3/2}}\right>.
\ee

Therefore if the condition 
\be
I(\tau) \equiv
\left< {d^2 \over d\tau^2} \sum_{i=1}^N\theta_i^2 \right> / ~(NT)
\ll 1,
\label{vircon}
\ee
is realized, then the virial relation 
\be
T\approx {1 \over 2N}
  \left< \sum_{i<j}^N  \frac{\ \theta_i \sin\theta_{ij}}
  {\sqrt{2}N(1-\cos\theta_{ij}+\epsilon)^{3/2}}\right>,
\label{virr}
\ee
holds. 

By using the total potential 
\be
V = 
 -\sum_{i<j}^N\frac{1}{\sqrt{2}N\sqrt{1-\cos\theta_{ij}+\epsilon}},
\label{pote}
\ee
the virial relation (\ref{virr}) can be expressed as
\be
T\approx 
  {1 \over 2N}\left< \sum_{i<j}^N \theta_i
  \frac{ \partial V}{\partial \theta_i}\right>.
\ee
Note that  even in the limit of $\epsilon \rightarrow 0$,
the virial relation
$\left< K \right>=2NT = -\left< V \right>/2$ is not justified,
because of the potential form (\ref{pote}). 
However,
as is shown in Sec.\ref{sec:SGR}, 
we can get the above ideal virial condition for 3-D gravity 
within the appropriate energy range.
 
\section{Virial condition in a \C-phase for SGR model}
\label{sec:viricon-core}

In \C-phase, 
almost all of the particles are trapped inside the \core, 
whose size is almost $\xi$. 
Therefore in this state canonical equations (\ref{canoeq3}) 
are approximated as
\bea
\frac{d\theta_i}{d\tau} 
  &=& p_i, \non \\
\frac{dp_i}{d\tau}      
  &=& -\sum_{j\neq i}^N
      \frac{\theta_{ij}}{2\sqrt{2}N\epsilon^{3/2}}
\non \\
  &=& -{\theta_i \over 2\sqrt{2}\epsilon^{3/2}}
      \left(1-\frac{1}{N}\right)
      +\sum_{j\neq i}^N
      \frac{\theta_{j}}{2\sqrt{2}N\epsilon^{3/2}},
\label{canoeq4}
\eea
since $|\theta_{ij}| \sim {\cal O}(\epsilon) \ll 1$.

In this approximation,
the second time derivative of the inertia moment of this system becomes
\bea
{d^2 \over d\tau^2} \sum_{i}\theta_i^2
  &=& 2\sum_i p_i^2-
      {1 \over \sqrt{2}\epsilon^{3/2}}
      \left(1-\frac{1}{N}\right)\sum_{i=1}^N \theta_i^2+
      \frac{2}{\sqrt{2}N\epsilon^{3/2}}
      \sum_{i<j}^N  \theta_{i}\theta_{j}.
\eea
Taking long-time average for this relation, we obtain 
\be
\left< {d^2 \over d\tau^2} \sum_{i}\theta_i^2 \right>
  = 4NT-{1 \over \sqrt{2}\epsilon^{3/2}}
    \left(1-\frac{1}{N}\right)\sum_{i=1}^N 
    \left< \theta_i^2\right>+
    \frac{2}{\sqrt{2}N\epsilon^{3/2}}
    \sum_{i<j}^N \left< \theta_{i}\theta_{j}\right>.
\label{appvir}
\ee
Since $\theta_i$ and $\theta_j (i \neq j)$ are 
independent stochastic variables, 
the last term of (\ref{appvir}) vanishes.
Hence from the virial relation,
\be
T \approx 
  {1 \over 4\sqrt{2}N N\epsilon^{3/2}}
  \left(1-\frac{1}{N}\right)
  \sum_{i=1}^N \left< \theta_i^2 \right>,
\label{virrB1}
\ee
and
\be
\left< K_{\rm p} \right> 
  \approx {G m^2 \over 2 \sqrt{2}\epsilon^{3/2}R}
           \left(1-\frac{1}{N}\right)
           \sum_{i=1}^N \left< \theta_i^2 \right>,
\label{virrB3}
\ee
hold. 
This shows that 
$\left< K_{\rm p} \right>$ in \C-phase is given by
the sum of $N$ statistical elements.


\end{document}